\documentclass[aps,physrev,preprint,groupedaddress]{revtex4-2}
\usepackage{amsmath}
\usepackage{amsfonts}
\usepackage{amssymb}
\usepackage{graphicx}
\usepackage{color}
\usepackage{subcaption}
\usepackage{setspace}
\usepackage{csquotes}
\usepackage[%
    colorlinks=true,
    pdfborder={0 0 0},
    linkcolor=red
]{hyperref}

\usepackage{array}

\begin{document}


\title{Impact of a Fano resonance on the measured transition time scale in solid state photoemission}

\author{Fei Guo$^{1,2}$}
\author{Dmitry Usanov$^{1}$}
\author{Eduardo B. Guedes$^{3}$}
\author{Arnaud Magrez$^{1}$}
\author{Michele Puppin$^{1,2}$}
\author{J. Hugo Dil$^{1,2,3}$}

\affiliation{
$^{1}$Institute of Physics, \'{E}cole Polytechnique F\'{e}d\'{e}rale de Lausanne, CH-1015 Lausanne, Switzerland\\ 
$^{2}$Lausanne Centre for Ultrafast Science (LACUS), \'{E}cole Polytechnique F\'{e}d\'{e}rale de Lausanne,
CH-1015 Lausanne, Switzerland\\
$^{3}$Center for Photon Science, Paul Scherrer Institut, CH-5232 Villigen, Switzerland
}

\date{\today}
\begin{abstract}
Fundamental quantum transition time scales are accessible through the spin polarization of photoelectrons coming from initially spin-degenerate states for solid-state materials . In this work we investigate the modification of this time scale in the vicinity of a Fano resonance in photoemission from a solid. We employ  spin- and angle-resolved photoemission spectroscopy (SARPES) to study the valence band of 1T-TiSe$_2$ and 1T-TiTe$_2$, with an excitation photon energy coinciding with the Ti 3p-3d autoionization state. The energy derivative of the measured spin polarization, which is in the off-resonance case proportional to the transition time, reveals a sign reversal and significant magnitude decrease compared to off-resonance measurements. We show that this effect goes beyond conventional semi-analytical models used to translate spin polarization to the EWS time delay. At the Fano resonance, the underlying interference assumption of the model breaks down, and additional information about resonance strength is needed to extract the transition time delays.
\end{abstract}

\maketitle
\section{Introduction}

In recent decades, chronoscopies of quantum processes has been a field of intensive research. Thanks to the advancement of ultrashort laser sources and the development of interference based measurement techniques, the attosecond dynamics of the photoemission process has been measured for atomic photoionization \cite{Eckle:2008, Schultze:2010, Klunder:2011, Neppl:2012} and solid-state photoemission from dispersive states \cite{Fanciulli:2017, Fanciulli:2017B, Guo:2025}, revealing time scales ranging from a few tens to a few hundreds of attoseconds. In the Eisenbud-Wigner-Smith (EWS) formalism, photoemission can be seen as a half-scattering process with a time scale determined by the scattering-induced phase shift of the outgoing wave packet \cite{Eisenbud:1948, Wigner:1955, Smith:1960}. The relative photoionization time scale between different states can be determined by streaking experiments, measuring in the time domain the peak of the photoelectron wave packet \cite{Cavalieri:2007,Neppl:2012, Lucchini:2015}; on the other hand, the phase of the wave packet can also be accessed via measurement of the spin polarization of photoelectrons emitted from spin-degenerate initial states\cite{Heinzmann:2012, Fanciulli:2017}. Given these experimental advancements, it is now possible to explore the possible factors which affect the attosecond transition time delay. For example, the influence of the initial state symmetry has been investigated for both atomic \cite{Ossiander:2017} and dispersive \cite{Guo:2025} systems. In this work, we aim at experimentally measuring the influence of a Fano resonances on the photoemission temporal dynamics in a solid. 

The Fano effect in photoemission originates from interference between photoelectrons and Auger-Meitner electrons emitted by core-hole relaxation in an autoionization process. The core-excited autoionization resonance is an example of a discrete state embedded in the continuum states, located at energies above the ionization threshold. The discrete-continuum interference gives rise to the characteristic Fano resonance line shape in the photoemission spectrum \cite{Fano:1961}.
The presence of such resonance is expected to induce dramatic changes on the outgoing electron wave packet, including its phase and thereby the EWS time delay, and could be used as a potential tuning knob of the microscopic ionization time, controlled by the excitation photon energy.
In the case of gas-phase photoionization, the attosecond photoionization dynamics was studied both theoretically \cite{Deshmukh:2018, Banerjee:2019} and experimentally \cite{Kotur:2016, Cirelli:2018} tuning the photon energy to characteristic energies, showing that in the vicinity of Fano resonances, the ionization intensity, phase and therefore EWS time delay vary significantly with energy and angle.

In the case of photoemission from solids, resonant photoemission (RPES) with photon energy corresponding to a core absorption edge has been extensively used as a way to investigate contribution of specific elements to the valence electronic structure \cite{Guillot:1977, Kay:1998, Sekiyama:2000, Fuentes:2001, Thomas:2003, Danzenbacher:2005, Krempasky:2016}. Past studies have also demonstrated that by using circular polarized light resonantly exciting a core level with a large spin-orbit splitting, along with a detailed analysis of the selection rules, one can distinguish between different spin initial states \cite{Tjeng:1997, Sinkovic:1997}. While this allows to characterize the specific autoionization channels for spin-orbit-split initial states, a study of the photoemission dynamics and phase shift at a generic Fano resonance has not been realized yet for solids. In this work, we present spin and angle-resolved photoemission spectroscopy (SARPES) results for the valence bands of 1T-TiSe$_2$ and 1T-TiTe$_2$ with photon energies resonant to the Ti 3p-3d autoionizing state, and we estimate the EWS time delay from the energy derivative of spin polarization with the aid of our model \cite{Fanciulli:2018} which translates spin polarization to photoemission phase shift. A large contrast in the energy derivative of the spin polarization is seen between on and off-resonance, which demonstrates the effect of multi-channel interference. It will be explained how the obtained results reflect the inappropriateness of certain assumptions of the analytical model \cite{Fanciulli:2018} in the presence of a resonance, and what extra information is needed to amend it.

\section{Methods}
1T-TiTe$_2$ and 1T-TiSe$_2$ single crystals were grown using the Chemical Vapor Transport (CVT) method with iodine (I$_2$) as the transport agent. High-purity Ti and Te or Ti and Se powders are sealed in a quartz ampoule and placed in a two-zone furnace with a controlled temperature gradient. For TiTe$_2$, the source is maintained at 800°C and the sink at 625°C, while for TiSe$_2$, the temperatures are slightly lower, with the source at 750°C and the sink at 600°C. After a few weeks, millimeter-sized crystals are obtained.

SARPES measurements were taken at the now decommissioned COPHEE end-station\cite{Hoesch:2002} on the SIS beamline of the Swiss Light Source, Paul Scherrer Institut. Band maps and spin polarization MDCs were measured with $\pi$-polarized light. Samples were cleaved at ultrahigh vacuum at T=25K, and the surface quality was checked by Low Energy Electron Diffraction (LEED). The measurement geometry is the same as that in ref.\cite{Fanciulli:2017}, in which the spin-resolved momentum distribution curves (MDCs) are obtained by rotating around the sample vertical axis. Spin asymmetry $A=\frac{N_{\uparrow}-N_{\downarrow}}{N_{\uparrow}+N_{\downarrow}}$ was measured for each spatial component ($x'$, $y'$, and $z'$) by two orthogonally positioned Mott detectors. Spin polarization was calculated by $P_i=\frac{1}{S}A_i$ where the Sherman function $S=0.08$. MDCs were measured for different binding energies in a randomized order to exclude sample aging effects. The $x$, $y$ and $z$ components of spin polarization mentioned in text are transformed to the sample frame \cite{Meier:2008}. The zero polarization on all MDCs was calibrated with respect to the general anti-symmetry of the spin texture, and the magnitude of spin polarization was calculated for a single, well-separated, peak for each material.

\section{Main}

\begin{figure}
    \centering
    \includegraphics[width=0.95\linewidth]{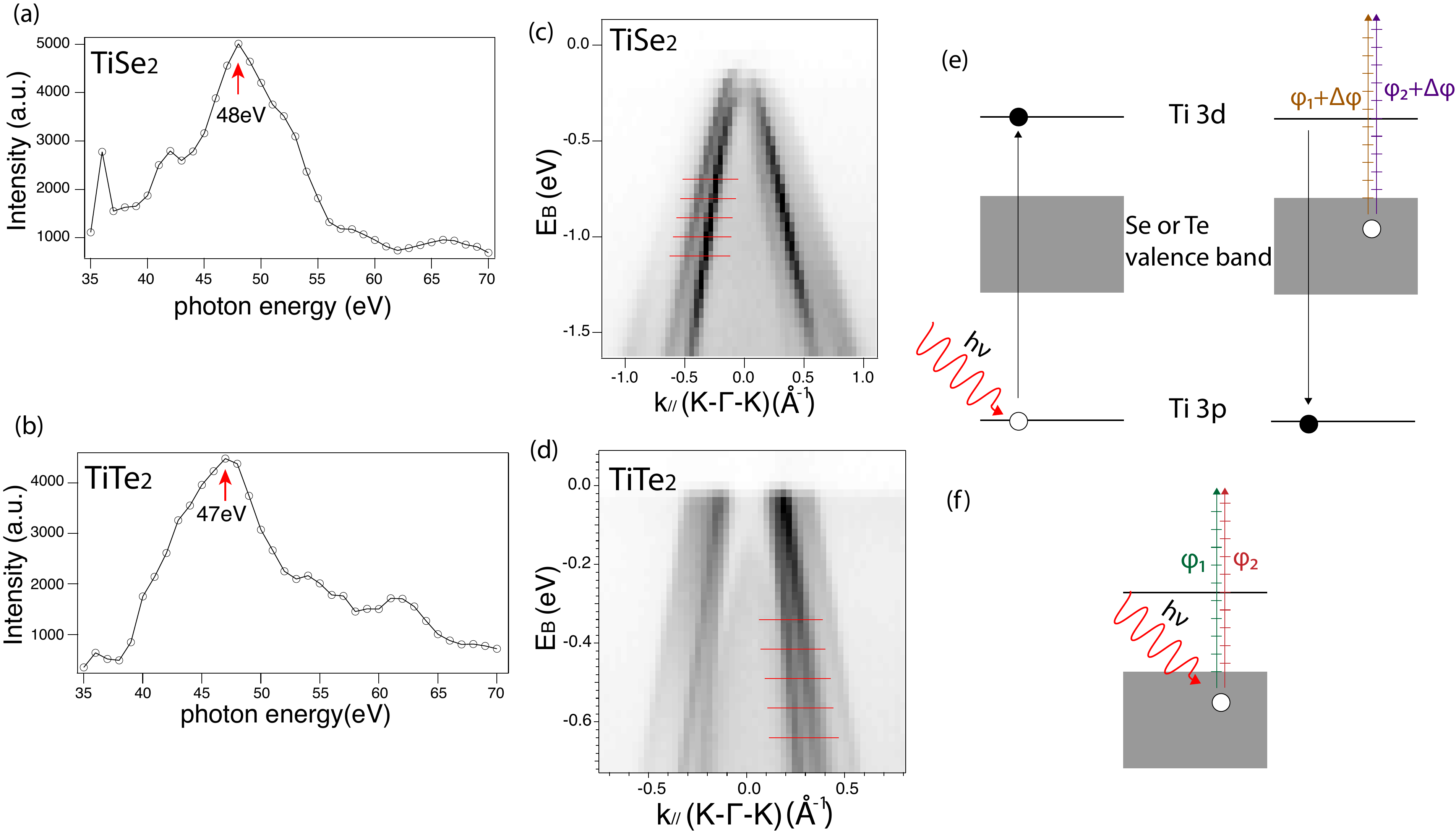}
    \caption{Photon energy scan of intensity at $E_b=0.5\,$eV for (a) TiSe$_2$ and (b) TiTe$_2$. Band maps of (c) Band map of TiSe$_2$ taken at $h\nu = 48\,$eV. (d) Band map of TiTe$_2$ taken at $h\nu = 47\,$eV, red lines indicate the binding energies at which momentum distribution curves (MDCs) were taken. Schematic illustrations of (e) Ti 3p-3d autoionization process and (f) direct photoemission process from the valence band.}
    \label{resonance1}
\end{figure}

TiSe$_2$ and TiTe$_2$ are both compounds of Ti, which has a $3p-3d$ absorption edge around 47\,eV \cite{Heise:1992, Thomas:2007}, well accessible by our experimental setup. The Ti 3p photo hole mainly decays by autoionization through Auger-Meitner (AM) electron emission. This therefore allows to investigate the influence of autoionization on the measured spin polarisation and the transition time scale. The process is illustrated schematically in Fig.\ref{resonance1}(e), when the excitation energy is tuned to this resonance, an electron in the Ti-3p state absorbs a photon and is excited to the Ti-3d conduction band, which exhibits strong hybridization with the Se-$4p_{x,y}$ or Te-$5p_{x,y}$ orbitals. This short-lived excited state decays by an AM process where a 3d electron decays back to the Ti-3p hole, while an electron is emitted from the valence band to conserve energy. This super-Coster-Kronig decay channel has the same final state as that of the direct photoemission from the valence band, Fig.\ref{resonance1}(f), and will therefore introduce an interference effect giving rise to a Fano line shape. 

As demonstrated by our model \cite{Fanciulli:2018}, multi-channel interference gives rise to spin polarization of photoelectrons originating from initially spin-degenerate states, and such spin polarization provides information about the transition phase, and thus the transition time scale. With this approach, the valence bands of both materials was investigated with SARPES at an off-resonance photon energy $h\nu = 67\,$eV, and the quantum transition time scale of photoemission was estimated from the energy derivative of photoelectron spin polarization. This resulted into $|\tau^s_{EWS}|\geq 152\,$as for TiSe$_2$ and $|\tau^s_{EWS}|\geq 142\,$as for 1T-TiTe$_2$ \cite{Guo:2025}. In this work, the experiments were repeated by matching the photon energy with the Ti $3p-3d$ absorption edge, thereby creating an additional resonant autoionization channel interfering with direct photoemission from the valence band.

To accurately locate the resonant excitation energy for each compound, a constant initial state spectrum was measured on the valence bands of both compounds at a binding energy $E_b=0.5\,$eV and around $k=0.25 \,\mathring{\text{A}}^{-1}$ in the $\Gamma-K$ direction, while scanning photon energy. The results are plotted in Figs.\ref{resonance1}(a) and (b) for TiSe$_2$ and TiTe$_2$ respectively. With a step size of 1\,eV, both spectra show a resonant behavior, and an intensity maximum appears at $h\nu = 48\,$eV for TiSe$_2$ and $h\nu = 47\,$eV for TiTe$_2$. Bandmaps were then taken at these photon energies, as shown in Figs.(\ref{resonance1})(c) and (d). SARPES measurements were performed for both compounds at the same 5 binding energies as the non-resonant study, which range from $E_b=0.7\,$eV to $1.1\,$eV for TiSe$_2$ as indicated in Fig.\ref{resonance1}(c), and from $E_b=0.34\,$eV to $0.64\,$eV for TiTe$_2$, as indicated in Fig.\ref{resonance1}(d).

\begin{figure}
    \centering
    \includegraphics[width=0.95\linewidth]{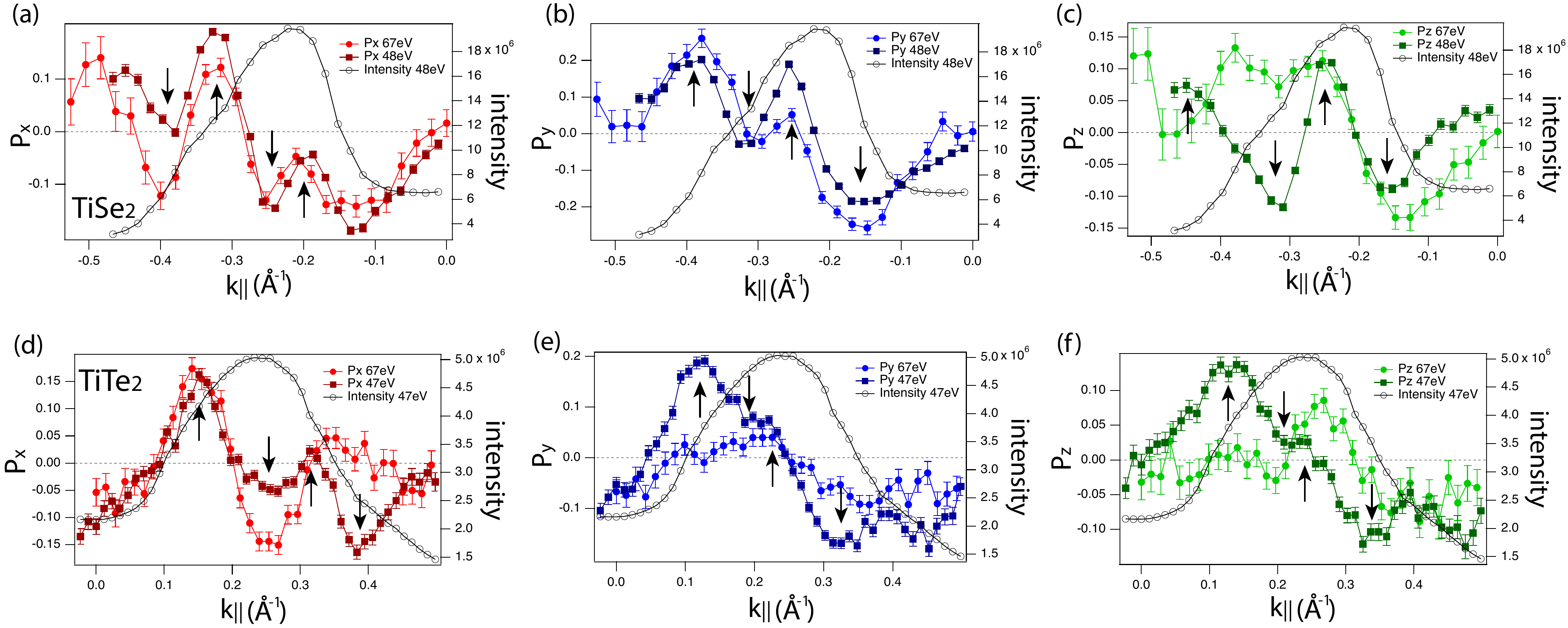}
    \caption{(a-c) Comparison of spin polarization MDCs of TiSe$_2$ at $E_b=0.9\,$eV in $x$, $y$ and $z$ directions, taken at $h\nu=67\,$eV and $h\nu=48\,$eV respectively. (d-f) Comparison of spin polarization MDCs of TiTe$_2$ at $E_b=0.49\,$eV in $x$, $y$ and $z$ directions, taken at $h\nu=67\,$eV and $h\nu=47\,$eV respectively.}
    \label{resonance2}
\end{figure}

\begin{figure}
    \centering
    \includegraphics[width=0.95\linewidth]{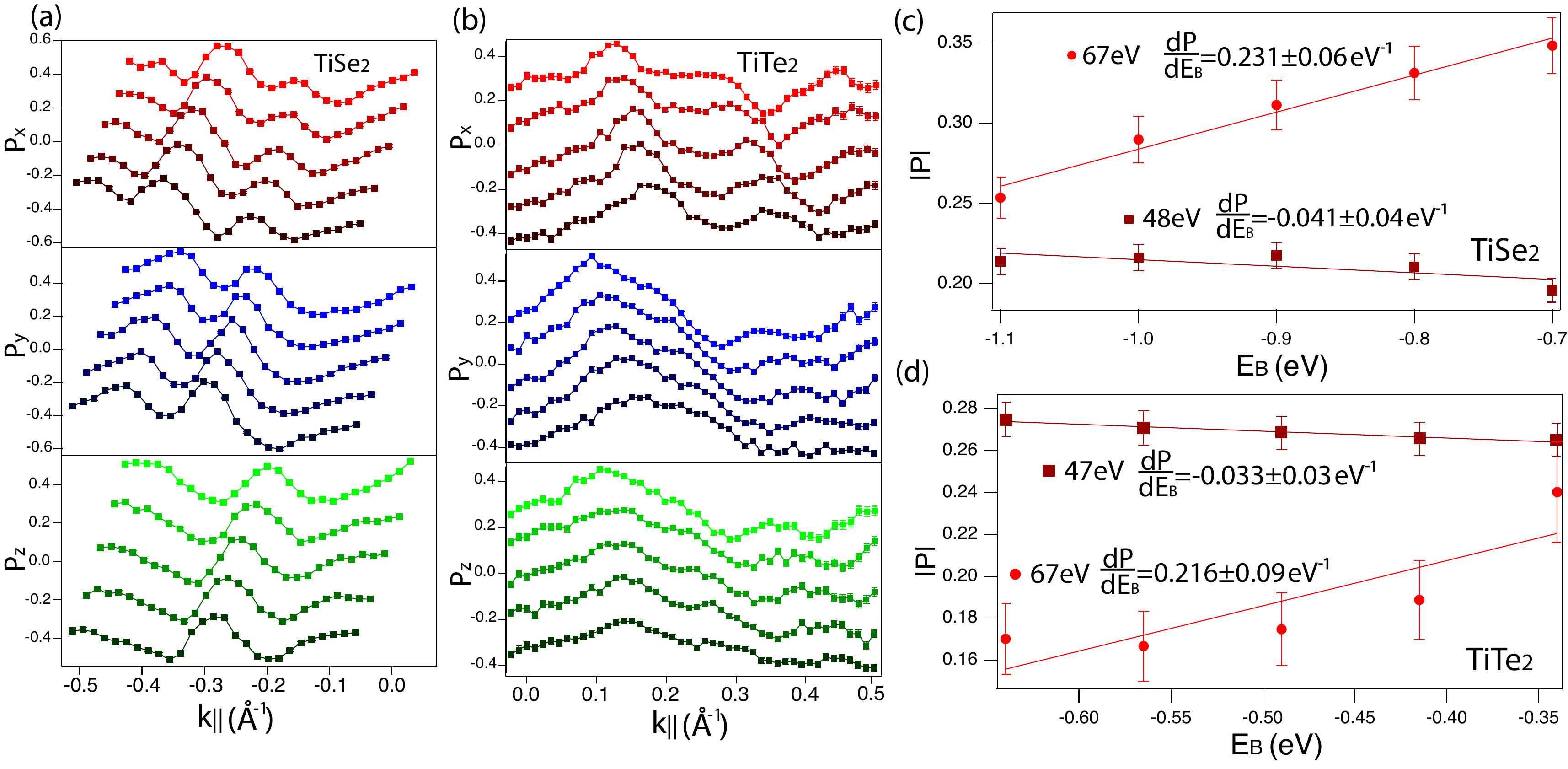}
    \caption{(a) Spin polarization MDCs of TiSe$_2$ with binding energies from top to bottom curves: $E_b=0.7\,$eV, 0.8\,eV, 0.9\,eV, 1.0\,eV, 1.1\,eV, in $x$ (red), $y$ (blue) and $z$ (green) directions, taken at $h\nu=48\,$eV, offset by 0.4, 0.2, 0, -0.2 and -0.4 respectively. (b) Spin polarization MDCs of TiTe$_2$ with binding energies from top to bottom curves: $E_b=0.34\,$eV, 0.415\,eV, 0.49\,eV, 0.565\,eV, 0.64\,eV, in $x$, $y$ and $z$ directions, taken at $h\nu=47\,$eV, offset by 0.3, 0.15, 0, -0.15 and -0.3 respectively. Maximum spin polarization magnitude at 5 binding energies with on and off-resonance photon energies, for (c) TiSe$_2$ and (d) TiTe$_2$, resulting estimates of on-resonance transition time scales close to 0.}
    \label{resonance3}
\end{figure}

At each binding energy, the momentum distribution curves (MDCs) of spin polarization were collected for the $x$, $y$ and $z$ directions, and are shown in Figs.\ref{resonance2}(a-c) for TiSe$_2$ at $E_b=0.9\,$eV and Figs.\ref{resonance2}(d-f) for TiTe$_2$ $E_b=0.49\,$eV, as a direct comparison with data collected at $h\nu=67\,$eV. On each MDC taken on resonance, two pairs of sign reversal features can be observed, marked by arrows in Fig.\ref{resonance2}. These features were also observed in previous studies \cite{Fanciulli:2017, Fanciulli:2017B, Fanciulli:thesis, Guo:2025} and were demonstrated to be a signature of multi-channel interference, and are hereafter referred to as the spin double polarization features (DPFs). For TiSe$_2$, the spin polarization has generally similar structures on and off-resonance, while the DPFs are more pronounced on resonance. For TiTe$_2$ on the other hand, the spin polarization in the $y$ and $z$ directions become stronger on resonance, showing clear DPFs which are less distinct at $h\nu = 67\,$eV. Spin polarization MDCs in all directions taken at all five binding energies are summarized in Fig.\ref{resonance3}(a) and (b) for each compound.

The maximum magnitude of \textit{\textbf{P}} for TiSe$_2$ was calculated from all three spatial components for the peak around $k=-0.32 \,\mathring{\text{A}}^{-1}$ in Figs.(\ref{resonance2})(a-c) at the five indicated binding energies as summarized in Fig.(\ref{resonance3})(c), with direct comparison to those extracted from MDCs taken at $h\nu = 67\,$eV. For TiTe$_2$ the same was done for the peak around $k=0.13 \,\mathring{\text{A}}^{-1}$ at the 5 indicated binding energies, as summarized in Fig.(\ref{resonance3})(d). As clearly visible from Figs. (\ref{resonance3})(c) and (d), the energy derivative of spin polarization $\frac{dP}{dE}$ reverses sign and decreases substantially in magnitude at resonant photon energies, and with the transition time scale proportional to $\frac{dP}{dE}$, it suggests estimated time scales close to 0.

This large contrast between on and off-resonance measurements certainly signals a change of the photoemission process, and it is interesting to identify what factors give rise to the significantly smaller values of $\frac{dP}{dE}$, and what are the implications about the photoemission phase shift on resonance. In the case of atomic photoionization, a study on valence electrons of argon showed that, in an energy region with a rich number of resonances, calculation of the wave packet group delay fails to reproduce experimental results \cite{Sabbar:2015}. This effect was interpreted in analogy to quantum tunneling. Since a quantum barrier and a resonance both act as an energy filter which reshapes the electron wave packet, the peaks of the incoming and outgoing wave packets may not correspond, and the wave packet group delay is not anymore a good representation of the ionization time delay. Although in case of the current experiment the synchrotron radiation bandwidth is much narrower than that of the high harmonic light source used for attosecond streaking, the photon energy giving maximum photoemission intensity was determined with a precision of 1\,eV, which may be a few hundreds of meV away from the actual maximum. Moreover, the resonant excitation energy is expected to lie somewhere between the maximum and minimum of the Fano profile \cite{Miroshnichenko:2010}. For photoemission from dispersive states, the asymmetric Fano profile cannot be resolved easily, and it is even more difficult to locate exactly the resonant photon energy. Therefore, the filter effect might persist in this experiment, and it is indeed possible that the `aggregate phase shift' deduced from spin polarization, is not an accurate reflection of the actual phase shift experienced during the half-scattering process.

In the context of our model, which considers two prominent interfering channels of similar weight, the spin polarization of photoelectrons can be written as a function of their relative phase \cite{Fanciulli:2018}:
\begin{equation}
    P\propto Im[M_1 \cdot M^*_2],
\label{P}
\end{equation}
and $\tau^s_{EWS}$ is approximated as a function of the relative phase shift $\phi_s$ of the 2 channels:
\begin{equation}
    \tau^s_{EWS}\propto\frac{d \sin(\phi_s)}{dE}\propto \frac{d P}{d E}.
\label{tau}
\end{equation}
However, in the more general case with more than 2 channels involved, the EWS time delay is a weighted sum accounting for all channels \cite{Pazuorek:2015}:

\begin{equation}
    \tau^s_{EWS}\approx\frac{\sum_{q}\hbar\frac{d \angle{M^q_{fi}}}{d E} |M^q_{fi}|^2}{\sum_{q}|M^q_{fi}|^2}.
\label{sum}    
\end{equation}
In the more complex scenario which cannot be simplified with two `virtual' channels, one needs to consider Eq.(\ref{sum}) in full and Eqs.(\ref{P}) and (\ref{tau}) lose their applicability. In the resonant photoemission case, the direct photoemission process and the autoionization process interfere to give rise to the Fano resonance, and each process is generally comprised of at least two channels. As the autoionization process has the same initial and final states as direct photoemission, the kinetic energy dependent phases of its two channels would also be the same $\varphi_1(E_k)$ and $\varphi_2(E_k)$, and an offset accounts for the photon energy dependent part, $\Delta \varphi(h\nu)$, as illustrated in Fig.(\ref{resonance1})(e) and (f). The photon energy dependent phase $\Delta \varphi(h\nu)$ varies significantly around a resonance \cite{Banerjee:2019, Barreau:2019}, but our measurement approach with constant photon energy is only sensitive to the kinetic energy dependent parts of the phase. If we consider two `virtual' channels representing autoionization and direct photoemission respectively, their phase difference is then only a function of photon energy, but not of kinetic energy. As evident from Figs.(\ref{resonance3})(c) and (d), it is indeed the case that the measured \textit{P} or $\sin(\phi_s)$ is nearly constant as a function of kinetic energy. Therefore, in order to get a meaningful estimate of $\tau^s_{EWS}$, one would need to consider all the channels independently. An interesting next step would be to repeat the measurement at several different photon energies, all in the vicinity of the Fano resonance with a spacing of a few hundreds of meV. Observing how spin polarization varies could potentially help to identify the variation of phase with photon energy $\Delta \varphi(h\nu)$. Moreover, if measurements are performed with increasing $h\nu$ away from the resonance, an abrupt change in $\frac{dP}{dE}$ would indicate the onset of the resonant photoemission regime. However, this would be a technically challenging experiment in the aspect of photon resolution, as well as in terms of time consumption.


\section{Conclusion}
In summary, valence bands of 1T-TiTe$_2$ and 1T-TiSe$_2$ have been investigated with SARPES as an attempt to estimate the EWS photoemission time delay. By tuning the excitation photon energy to the Ti 3p-3d Fano resonance, we demonstrate a significant decrease in the measured energy derivative of spin polarization, compared to that obtained with off-resonance photon energy. This decrease in $\frac{dP}{dE}$ is a consequence of strong interference between direct photoemission and autoionization processes, which are both generally multi-channel. In the framework of our model, describing each process as a `virtual' channel results in a relative phase between the two `virtual' channels that is only a function of photon energy, and is no longer an adequate representation of the phase shift induced by photoemission. A proper estimate of the resonant EWS time delay requires full characterization of all ionization channels, and this requires repeating SARPES measurements in a narrow range of photon energies around the resonance, which is highly demanding in photon energy precision, photon flux, and measurement time. On the other hand, direct time delay measurement with attosecond light sources for solids is also challenging as it compromises energy and angle resolution, and matching the high harmonic photon energy with an absorption threshold might only be possible for some specific elements. Although significant experiment improvement is needed for estimating the resonant EWS time delay for solids, our work gives an insight into the role of additional ionization channels in the measured spin and phase shift, and opens up a field for development of new approaches.

\section*{Acknowledgement}

F.G. and J.H.D. acknowledge support from the Swiss National Science Foundation (SNSF) Project No. 200021-200362. D.U. acknowledges support from the Swiss National Science Foundation through the Spark grant CRSK-2\textunderscore228962

\bibliographystyle{apsrev4-1}
\bibliography{References_SOIS}

\end{document}